\input harvmac

\Title{\vbox{\baselineskip12pt
\hbox{CERN-TH/97-94}
\hbox{CPTH-S512.0697}	
\hbox{hep-th/9707013}}}
{\vbox{\centerline{$R^4$ Couplings in $M$ and Type $II$ Theories}
\vskip2pt
\centerline{on Calabi-Yau spaces}}}

\centerline{I. Antoniadis$^{a,b}$, S. Ferrara$^{a}$,
R. Minasian$^{a}$ and K.S. Narain$^{c}$} 
\bigskip\centerline{$^a${\it CERN, CH-1211 Geneva 23, Switzerland}}
\medskip\centerline{$^b${\it Centre de Physique Th\'eorique, 
Ecole Polytechnique,\footnote{$\dagger$}{Laboratoire Propre du CNRS UPR
A.0014.} 91128 Palaiseau, France}}
\medskip\centerline{$^c${\it International Centre for Theoretical Physics, 
34100 Trieste, Italy}}
\vskip .3in

We discuss several implications of $R^4$ couplings in $M$ theory when
compactified on Calabi-Yau (CY) manifolds. In particular, these couplings can
be predicted by supersymmetry from the mixed gauge-gravitational
Chern-Simons couplings in five dimensions and are related to the one-loop
holomorphic anomaly in four-dimensional $N=2$ theories. We find a new
contribution to the Einstein term in five dimensions proportional to the
Euler number of the internal CY threefold, which corresponds to a one-loop
correction of the hypermultiplet geometry. This correction is reproduced
by a direct computation in type $II$ string theories. Finally, we discuss a
universal non-perturbative correction to the type $IIB$ hyper-metric.

\vskip .3in
\Date{\vbox{\baselineskip12pt
\hbox{CERN-TH/97-94}
\hbox{June 1997}}}

\def\sqr#1#2{{\vbox{\hrule height.#2pt\hbox{\vrule width
.#2pt height#1pt \kern#1pt\vrule width.#2pt}\hrule height.#2pt}}}
\def\Box{\mathchoice\sqr64\sqr64\sqr{4.2}3\sqr33}

\def\ie{{\it i.e.,}\ }

\lref\dufflm{M.J.~Duff, J.T.~Liu and R.~Minasian, Nucl. Phys. 
{\bf B452} (1995) 261.}

\lref\witfb{E.~Witten,  Nucl. Phys. {\bf B463} (1996) 383.}

\lref\duffmw{M.~J.~Duff, R.~Minasian and E.~Witten, hep-th/9601036.}

\lref\hult{C.~M.~Hull and P.K.~Townsend, Nucl. Phys. {\bf B438} (1995) 109.}

\lref\wit{E.~Witten, Nucl. Phys. {\bf B443} (1995) 85.}

\lref\schwarz{J.H.~Schwarz, hep-th/9510086; hep-th/9601077.}

\lref\horw{P.~Ho\v rava and E.~Witten, Nucl. Phys. {\bf B460} (1996) 506.}

\lref\witsi{E.~Witten, Nucl. Phys. {\bf B460} (1996) 541.}

\lref\antft{I.~Antoniadis, S.~Ferrara and T.R.~Taylor, Nucl. Phys.
{\bf B460} (1996) 489.}

\lref\becker{K.~Becker, M.~Becker and A.~Strominger, Nucl. Phys.
{\bf B456} (1995) 130.}

\lref\cadcdf{A.C.~Cadavid, A.~Ceresole, R.~D'Auria and S.~Ferrara,
Phys. Lett. {\bf B357} (1995) 76.}

\lref\gst{M.~G\"{u}naydin, G.~Sierra and P.K.~Townsend, Nucl. Phys. 
{\bf 242} (1984) 244.}

\lref\cardoso{G.~L.~Cardoso, G.~Curio, D.~L\"{u}st, T.~Mohaupt and S.-J.~Rey,
hep-th/9512129.}

\lref\kap{V.~Kaplunovsky, J.~Louis and S.~Theisen, Phys. Lett.
{\bf B357} (1995) 71.}

\lref\hly{S.~Hosono, B.~H.~Lian and S.~T.~Yau, alg-geom/9511001.}

\lref\town{P.~K.~Townsend, Phys. Lett. {\bf B373} (1996) 68.}

\lref\papa{G.~Papadopoulos and P.~K.~Townsend, Phys. Lett. {\bf B357} (1995) 
300.}

\lref\hosono{S.~Hosono, A.~Klemm, S.~Theisen and S.~T.~Yau, 
Comm. Math. Phys. {\bf 167} (1995) 301.}

\lref\kachru{S.~Kachru and C.~Vafa, Nucl. Phys. {\bf B450} (1995) 69.}

\lref\mirror{S.~Ferrara, J.A.~Harvey, A.~Strominger and C.~Vafa,
Phys. Lett. {\bf B361} (1995) 59.}

\lref\bers{M.~Bershadsky, S.~Cecotti, H.~Ooguri and C.~Vafa,
Comm. Math. Phys. {\bf 165} (1994) 311.}

\lref\fff{M.~Bershadsky, S.~Cecotti, H.~Ooguri and C.~Vafa,
Nucl. Phys. {\bf B405} (1993) 279.}

\lref\kkoun{E.~Kiritsis and C.~Kounnas, 
Nucl. Phys. {\bf B442} (1995) 472.}

\lref\agnt{I.~Antoniadis, E.~Gava, K.S.~Narain and T.R.~Taylor, 
Nucl. Phys. {\bf B407} (1993) 706.}

\lref\agnta{I.~Antoniadis, E.~Gava, K.S.~Narain and T.R.~Taylor, 
Nucl. Phys. {\bf B413} (1994) 162.}

\lref\klemm{A.~Klemm, W.~Lerche and P.~Mayr, Phys. Lett.
{\bf B357} (1995) 313.}

\lref\vafawit{C.~Vafa and E. Witten, Nucl. Phys. {\bf B447} (1995) 261.}

\lref\lnsw{W. Lerche, B.E.W. Nilsson, A.N. Schellekens, N.P. Warner, Nucl.
Phys. {\bf B299} (1988) 136.}

\lref\ipw{ C.J. Isham, C.N. Pope and N.P. Warner, Class. Quant. Grav. 
{\bf 5} (1988) 1297.}

\lref\gvvz{M.T. Grisaru, A.E.M. van de Ven and D. Zanon,
Nucl. Phys. {\bf B277} (1986) 409.}

\lref\ats{ A.A. Tseytlin, Nucl. Phys. {\bf B467} (1996) 383.}
 
\lref\gmg{M.B. Green and M.  Gutperle, hep-th/9701093.}

\lref\gvh{M.B. Green and P. Vanhove, hep-th/9704145.}

\lref\bb{ K. Becker and M. Becker, Nucl. Phys. {\bf B477} (1996)
155.}

\lref\svw{ S. Sethi, C. Vafa and E. Witten, Nucl. Phys. {\bf B480} (1996)
213.}

\lref\chs{ C.G. Callan, J.A. Harvey and A. Strominger, Nucl. Phys. {\bf
B467} (1991) 60.}
 
\lref\witcy{E.~Witten, hep-th/9602070.}

\lref\fkm{S. Ferrara,  R.R. Khuri and R. Minasian, Phys. Lett.
{\bf B375} (1996) 81.}

\lref\grwit{D.J. Gross and E. Witten, Nucl. Phys. {\bf B277} (1986) 1.}

\lref\frpope{M.D. Freeman and C.N. Pope, Phys. Lett.
{\bf B174} (1986) 48.}

\lref\mikerev{M.J. Duff, Int. J. Mod. Phys. {\bf A11} (1996) 5623;
C. Vafa, hep-th/9702201.}

\lref\vafarev{C. Vafa, hep-th/9702201.}

\lref\schrev{J.H. Schwarz,  hep-th/9607067.}

\lref\cfg{S. Cecotti, S. Ferrara and L. Girardello, Int. J. Mod. Phys.
{\bf A4} (1989) 2475.}

\lref\fsab{S. Ferrara and S. Sabharwal,  Nucl. Phys. {\bf B332} (1990)
317.}

\lref\cand{P. Candelas, X.C. De La Ossa, P.S. Green  and L. Parkes,
Nucl. Phys. {\bf B359} (1991) 21.}

\lref\ooguri{H. Ooguri and C. Vafa, Phys. Rev. Lett. {\bf 77} (1996)
3296.}

\lref\cdp{P. Candelas, E. Derrick and L. Parkes, Nucl. Phys. {\bf B407}
(1993) 115.}

\lref\fhsv{S.~Ferrara, J.A.~Harvey, A.~Strominger and C.~Vafa,
 Phys. Lett. {\bf B361} (1995) 59.}
 
\lref\hmoore{J.A.~Harvey and G. Moore, hep-th/9610237.}
 
\lref\townss{P.~K.~Townsend, hep-th/9507048.} 

\lref\bars{I. Bars and S. Yankielowicz, Phys. Rev. {\bf D53} (1996) 4489; 
I. Bars, Phys. Rev. {\bf D55} (1997) 3633.}


\newsec{Introduction}

It has recently been  shown that certain $R^4$ terms, present in type $IIA$
and type $IIB$ \gmg\ theories, suggest the existence of similar terms in
$M$ theory \gvh. In \gvh\ a possible relation to the anomaly-cancelling
$A_3 \wedge I_8(R)$ term \dufflm\ due to supersymmetry was argued. In this
note, we present further evidence of these terms looking
at compactifications of $M$ theory to lower odd-dimensional theories on
Calabi-Yau manifolds. These compactifications are known to relate $M$
theory to several dual partners, the most explored examples of the
dualities being nine-dimensional duality between $M$ theory on $T^2$ and
type $IIB$ on $S^1$ \schrev\ and between $M$ theory on $X_n$ and
heterotic string on $X_{n-1} \times S^1$ in dimensions $11-2n$, where
$X_n$ is a CY manifold of complex dimension $n$ (for recent
reviews, see \mikerev).

In $D=9$ a one-loop calculation of the anomaly-generating function \lnsw\
gives a result similar to the one of the ten-dimensional type $IIA$
calculation \vafawit\ and is of the form $\int A_1 \wedge I_8(R),$ where 
$I_8(R)$ is  an eight-form polynomial in curvature \dufflm, and the vector
can be identified with $B_{\mu 9}$ in type $IIA$ and with the Kaluza-Klein 
gauge boson
$g_{\mu 9}$ in type $IIB$.  While in $M$ theory this term is present
already in eleven dimensions, ten-dimensional type $IIB$ theory forbids a
similar term  \vafawit. However, its presence  is supported
by the $T$-duality between these two theories and is consistent with
$SL(2, Z)$ duality in nine dimensions. This will be discussed in detail
in Section 4, where we also present a complementary argument and show the
emergence of a chiral fivebrane in the nine-dimensional type $IIB$ spectrum.

In $D=7$, $R^4$ coupling on $K3$ yields an $R^2$ coupling needed to
reproduce the dilaton equation of motion of heterotic string on $T3$
\eqn\dila{\Box \phi_7 = R^2 + \ldots ,}
that is the supersymmetric extension of the string Bianchi identity. The
heterotic dilaton $\phi_7$ is given by the $K3$ volume,
$e^{-2\phi_7}=V_{K_3}$ \refs{\vafawit, \dufflm, \hmoore}.

The $D=5$ case will be discussed in detail in Section 2. We will
show that the $R^4$ terms generate two terms. One is the superpartner of
the gravitational Chern-Simons term and is of the form
\eqn\supercs{\sum_{h_{(1,1)}}\alpha_{\Lambda} \int t^{\Lambda}  R\wedge
R\wedge e,}
where $t^{\Lambda}$ are five-dimensional  special coordinates
(subject to constraint ${1 \over 6} t^{\Lambda} t^{\Delta} t^{\Sigma}
C_{\Lambda
\Delta \Sigma}$ = 1), and the constants $\alpha_\Lambda$ will be defined
in the next section.
By further compactification to four dimensions, the two combine into
\eqn\fourcompl{\sum_{h_{(1,1)}}\alpha_{\Lambda} \int Z^{\Lambda}  R^2,}
where $Z^{\Lambda} = {A_5}^{\Lambda} + i t^{\Lambda} r_5$, $r_5$ being the
radius of the fifth dimension and ${A_{\mu}}^{\Lambda}$
the five-dimensional gauge fields. 
The other remnant of $R^4$ in $D=5$ is in the two-derivative part of the
effective field theory; it is proportional to the Euler number
and can be regarded as a  correction to the hypermatter
geometry. We reproduce this correction by a direct one-loop string
computation in $D=4$, in Section 3.

In $D=3$, we find a cosmological constant proportional to the
Euler number of the internal fourfold, which is the supersymmetric extension
of the tadpole term $\chi A_3$.

{}Finally, in Section 5, we discuss type $IIB$ compactifications to 
$D=4$ on Calabi-Yau. In particular, we extract a universal non-perturbative
correction to the hypermultiplet metric, which exhibits $SL(2,Z)$ invariance,
and speculate about its possible extension to a quaternionic symmetry.

\newsec{$M$ Theory and type $IIA$ theory on Calabi-Yau Threefolds}

The bosonic action of the eleven-dimensional supergravity limit of
$M$ theory
is given by
\eqn\elevsg{I_{11}={1\over 2} \int_{M^{11}} d^{11}x 
\left[\sqrt{-g}R - {1 \over 2} F_4 \wedge *F_4 - {1 \over 6} A_3 
\wedge F_4 \wedge F_4\right].}
This action should be implemented by a term predicted by membrane/fivebrane
duality \dufflm. Indeed, cancelling the anomaly on the fivebrane
worldvolume by a bulk contribution determines a coupling  between   the
three-form potential and an eight-form  polynomial in curvature.
\eqn\newel{I_{11}^{Lorentz} = \int_{M^{11}} A_3\wedge
{1\over (2\pi)^4}\left[-{1\over 768}(\tr R^2)^2+{1\over 192}\tr R^4\right].}
The gravitational constant and the membrane and
fivebrane tensions are set to one. 

The reductions of the effective theory on Calabi-Yau manifolds have been
extensively discussed during the last year. In particular, in five
dimensions
it is well known 
(see, {\it e.g.}, \refs{\cadcdf,\antft}) that in addition to $h_{(1,1)}$
vectors and $h_{(2,1)}+1$ hypermultiplets, the theory has a geometrical
coupling term \gst
\eqn\topcop{I_5=- {1 \over 12}C_{\Lambda\Sigma\Delta}\int_{M^5} 
A_1^\Lambda \wedge F_2^\Sigma \wedge F_2^\Delta.}
The $U(1)$ fields are normalized so that 
they couple to integer charges. On the other hand, the reduction of
\newel\ yields an interaction of the form
\eqn\new{I_5^{Lorentz} \sim \int_{M^5} \alpha_{\Lambda} A^\Lambda_1 \wedge 
\tr R^2.} 
As discussed in \fkm, \new\ can be viewed as a bulk term needed
for cancelling the anomalies due to the wrapping of the fivebrane on the
CY four-cycles.
The $\alpha_\Lambda$ define the topological couplings:
\eqn\alp{\alpha_\Lambda={1\over 16(2\pi)^2}\int_{X_6} \omega_\Lambda \wedge 
\tr R^2,}
where $\Lambda=1,...,h_{(1,1)}$  and $\omega_\Lambda$ is the corresponding
$(1,1)$ harmonic form. 

In analogy with the lower-dimensional Green-Schwarz terms, we claim that
in eleven dimensions not only \newel\ should be added to the action but
also its ``supersymmetric'' partner (the reference to supersymmetry is
rather indirect here since it will be argued not directly in the
eleven-dimensional theory but after its compactification):
\eqn\newrr{{\hat t}^{\mu_1 \cdots \mu_8 }  {\hat t}_{\nu_1 \cdots \nu_8}
R_{\mu_1
\mu_2}^{\nu_1 \nu_2}  \cdots  R_{\mu_7 \mu_8}^{\nu_7 \nu_8},}
where for any antisymmetric matrix $M$, ${\hat t}^{\mu_1 \cdots \mu_8 }
M_{\mu_1  \mu_2} \cdots  M_{\mu_7 \mu_8}\equiv 24 tr M^4 - 6 (tr M^2)^2 +1/2
\epsilon^{\mu_1 \cdots \mu_8 } M_{\mu_1 \mu_2} \cdots  M_{\mu_7
\mu_8}\equiv t_8 M^4 +{1 \over 2} \epsilon \cdot M^4.$  \foot{We follow the
conventions of \refs{\ats,\gmg,\gvh}. In particular, we
take the normalization of the eleven-dimensional term with two $t_8$ to be
${\pi^2 \over 9 \cdot 2^7} \int d^{11}x  \sqrt{-g} t_8 t_8 R^4. $
Moreover $\epsilon$ denotes the totally antisymmetric tensor in 8 dimensions
with Lorenzian signature, throughout the paper.}
Clearly, \newrr\ has a parity-violating piece, containing, one $\epsilon$
tensor, and this is exactly the piece contributing to \newel. The
condition of existence of  nowhere-vanishing spinors on the background
imposes further constraints \ipw\ and the integral over \newrr\ can be
replaced by an integral over a parity-preserving combination, 
$t_8 t_8 R^4  - {1 \over 4} \epsilon \epsilon R^4,$ which also appears
in the expression of the four-loop $\sigma$-model beta-function \gvvz, 
up to the relative sign which we discuss below. This is what we call
the $R^4$ term in the eleven-dimensional effective theory. 
By simple scaling argument, it can be shown that this reduces to a
one-loop $R^4$ term in type $IIA$  theory.

As mentioned before, we would like to argue that the compactifications of
this term yield results in complete agreement with supersymmetry. Let us
concentrate on the $D=5$ $N=1$ case obtained by compactification on a
Calabi-Yau manifold. As was pointed out in ref. \fkm, here (in the
decompactification
limit of $D=4$ $N=2$) all the instanton corrections are suppressed and we
see the leading term in the (vector-valued) holomorphic coupling {\new}. The
reduction of \newrr\ will give the supersymmetric completion of this
coupling, but also we will see a new  effect in the
universal hypermultiplet. Before we perform the reduction, we notice 
that besides the low derivative $R^2$ and $R$ terms, the
eleven-dimensional $R^4$ coupling is
going to give rise also to a similar higher-derivative term in $D=5$, 
which we do not discuss here. 

Let us first write $R^4$ in $D=11$ in a more convenient form for our purposes:
\eqn\newrnew{ Y_{11} \equiv {-\pi^2 \over 32}  \left[4 \int R\wedge R\wedge
R\wedge R\wedge e\wedge e\wedge e - \int \sqrt{-g} R \cdot R \cdot R \cdot R
\right],}
where $R \cdot R \cdot R \cdot R = 6t_8(4trR^4 - (trR^2)^2) = 12
(R_{\mu \nu \rho \lambda}R^{\mu \nu \rho \lambda})^2 +  \ldots$. We have
written explicitly only the term essential for the
threefold compactifications. On $M_{11}
= M_5 \times X_6$  there are two contributions from \newrnew.
The first, moduli dependent, reproduces the form familiar from the
one-loop string formula for the gravitational holomorphic coupling
\eqn\couplone{ {-1 \over 8}(c_2 \cdot \vec J)  R\wedge R\wedge e,}
where the internal part is
\eqn\oner{\int_{X_6} R\wedge R\wedge e\wedge e = {1 \over (2 \pi)^2} (c_2
\cdot \vec J).}

Equation \couplone\ is in fact the supersymmetric partner of \new. Upon
reduction to $D=4$, these two terms form the coupling $\alpha_{\Lambda}
Z^{\Lambda} R^2$ where $Z$'s are
the complex fields in $N=2$ vector multiplets.
Note that \couplone\ receives a contribution from the fully
contracted combination of four $R$'s. This contribution is exactly equal
to the one from the wedged product, and this turns out to be an extremely
important consistency check. Referring to \gmg\ for more detailed
discussion
of this point, we just mention that the relative minus sign is fixed in
Type $IIA$, while in type $IIB$ the sign is ambiguous. As will be
shown in Section 6, $N=2$ supersymmetry requires the four-dimensional
analogue
of \couplone\ to have a vanishing coefficient for type $IIB$ compactifications
which therefore fixes the sign ambiguity to be the opposite from that of type
$IIA$..

In ten dimensions, type $IIA$ has also a tree-level $R^4$ term which goes to
zero in the eleven-dimensional limit \gvh. The relative sign between
$t_8 t_8$  and $\epsilon \epsilon$ is the same at tree level 
in both type $IIA$ and $IIB$
theories, following the four-loop $\sigma$-model computation \gvvz.
It follows that
type $IIA$ has a relative sign flipped between tree-level and one-loop
terms, while type $IIB$ does not, consistently with $SL(2,Z)$ symmetry. Note that
because of this relative sign between $t_8 t_8$ and $\epsilon
\epsilon$ terms the reduction of the
tree-level terms to four dimensions gives a vanishing
contribution to $R^2$ for both type $IIA$ and $IIB$ as required by $N=2$
supersymmetry.

It is instructive to recall the $K3$ compactification and discuss the 
analogue of \newrnew\ in seven dimensions. 
In this case, \oner\ is  replaced by a constant equal to the Euler number of
$K_3$ and the $R^2$ coupling is simply the supersymmetric extension of 
the Chern-Simons gravitational couplings in $D=7$ proportional to the first
Pontryagin number discussed in \dufflm. Moreover, upon compactification to six
dimensions, it corresponds to a one-loop correction in the type $IIA$ theory
compactified on $K3$. Note that due to the sign flip between the $t_8 t_8$ and
$\epsilon \epsilon$ term for the tree-level type $IIA$ and $IIB$ and for the one
loop for $IIB$ there is no $R^2$ term at the tree level for both $IIA$ and $IIB$
and at the one loop level for $IIB$. The results for type $IIA$ on $K3$ are
consistent with what one expects from the 6-dimensional heterotic-type $IIA$
duality. Indeed it is easy to see that the type $IIA$ tree-level and one-loop
contributions to $R^2$ are mapped via duality to heterotic one-loop and tree-level 
contributions respectively and in the heterotic theory compactified on $T^4$  there
is a tree-level $R^2$ term but no one-loop term.

The second  term which arises after integration on CY is
moduli-independent and  is proportional to the Euler number of the
internal manifold:
\eqn\twor{\int_{X_6} R\wedge R\wedge R = {1 \over 3! \, (2 \pi)^3} \chi ,}
yielding to a correction to the Einstein term of the form  $\sqrt{-g} R
\chi$. This is not the whole story yet. There is a modification of the
eleven-dimensional Einstein equation due to the higher curvature term. On
$M^{11}= M^5 \times X_6$, the non-vanishing component is  \refs{\frpope, \grwit}
\eqn\einstmod{R_{i {\bar j}}=\partial_i \partial_{\bar j} X,}
where $X$ is the six-dimensional Euler integrand. This leads to a correction in
the kinetic terms of $h_{(1,1)}-1$ vector moduli.

The overall effect is a modification of the usual five-dimensional action by a 
shift in the universal hypermultiplet\foot{Our discussion throughout this
paper is applied only to the non-universal hypermultiplets, since we treat the
universal one as constant.} which contains the CY volume of $X_6$, 
${\cal V} = e^{-2 \phi_5}$ \cadcdf. In fact, the scalar kinetic terms 
(with the properly redefined special coordinates $t^{\Lambda}$)
in five dimensions in the $M$ theory frame should be obtained by reducing
$M$
theory on the Calabi-Yau space. The string calculation for type $IIA$
presented in the next section implies that these kinetic terms are of the
form:
\eqn\couplhype{\sqrt{-g}\left[ (e^{-2\phi_5} - 
{1 \over 12 {\pi}} \chi)
(R + G_{\Lambda \Delta} \partial t^{\Lambda} \partial t^{\Delta}) +
(e^{-2\phi_5}+ {1 \over 12 {\pi}} \chi)  G_{q \bar{q}}
\partial q    \partial{\bar q}
\right],} 
Here $q$ are the hypermultiplets and $t^{\Lambda}$ are the 5-dimensional 
special coordinates subject to the
constraint ${1\over 6}t^{\Lambda} t^{\Delta} t^{\Sigma} C_{\Lambda \Delta
\Sigma} =1$. The above constraint defines a $h_{(1,1)} -1$ dimensional
surface \refs{\gst, \cadcdf}, and the kinetic term for $t^{\Lambda}$ contains
the induced metric on this surface. 
A redefinition of the five-dimensional dilaton $\phi_5$ is translated into
a one-loop correction to the hypermultiplet metric
$G_{q {\bar q}}  \rightarrow  
G_{q {\bar q}}[ 1+ e^{2\phi_5}\chi /(6 \pi)].$ 
This one-loop correction in the hypermultiplet 
geometry will be confirmed in the next section, where we show that this is 
indeed the case by performing a direct string computation in four dimensions.
 
In fact going to $D=4$, $M$ theory compactification on CY$\times S^1$ is
believed to describe the strong coupling regime of
type $IIA$ on the same CY. In ten dimensions type $IIA$ theory has $R^4$ 
correction both at the tree level and at one loop level. As mentioned earlier 
the relative sign
between the $t_8 \cdot t_8$ term and $\epsilon \cdot \epsilon$ term is 
opposite between the tree
level and the one loop level. To be explicit, let us denote by $Y_0$ and
$Y_2$ respectively the $t_8 \cdot t_8$ and 
$\epsilon \cdot \epsilon$ parts 
of $R^4$ terms. Then the $R^4$ term in the ten dimensional 
action for type $IIA$ is obtained from the combination $(t_8
+{i\over 2}\epsilon) \cdot (t_8
-{i\over 2}\epsilon)$ for the tree level while from $(t_8\cdot t_8
-{1\over 4}\epsilon \cdot \epsilon)$ for one loop, namely: 

\eqn\IIAY{\int d^{10} x \sqrt{-g}[ \zeta(3)e^{-2\phi}(Y_0 + Y_2)+
(Y_0 - Y_2)]\ .}
In type $IIB$, as we shall discuss in Section 5, the
tree level and the one-loop terms will be identical. 
Equation of motion for the internal metric and the dilaton to this order gets
contribution only from $Y_0$ as discussed in \refs{\frpope, \grwit}. 
The corrected metric 
turns out to be still K\"ahler though not Ricci flat. Denoting by $\delta
g$ the correction to the Calabi-Yau metric one finds

\eqn\Rij{R_{i\bar{j}} = -{1\over2}\partial_i \partial_{\bar{j}}Tr\delta g
={1 
\over {3!
(2\pi)^3}}(2\zeta(3)+ e^{2\phi_0}{2\pi^{2} \over 3})\partial_i
\partial_{\bar{j}}X}
\eqn\dilat{\phi = \phi_0 + {1 \over {12 
(2\pi)^3}}(2\zeta(3)+ e^{2\phi_0}{2\pi^{2}\over 3})X}
where $\phi_0$ is the constant uncorrected dilaton.

The kinetic terms for scalars corresponding to the deformation of K\"ahler 
class and complex structures receive correction only from $Y_0$ term, as 
$Y_2$ will necessarily involve four 4-dimensional indices and hence four 
derivatives. It follows that the correction to the moduli metric is of the 
same form for tree level and one loop. Although we have not obtained the 
correction to the moduli metric directly by reducing the ten dimensional 
action on the corrected Calabi-Yau space, the string calculation presented 
in the next section implies the following correction to the kinetic terms 
(in the string frame):

\eqn\IIAmetric{-\int d^4x \sqrt{-g} {\chi \over {(2\pi)^3} }
(e^{-2\phi_4}
{2\zeta(3)
\over {\cal V}} + {2\pi^{2} \over 3}) (G_{\Lambda \Delta}\partial
Z^{\Lambda} \partial
{\bar{Z}}^{\Delta}-  G_{q\bar{q}}\partial q \partial{\bar{q}})}
where $q$ are the scalars of
hypermultiplets (\ie complex structure moduli) and
$Z$ are the scalars in the vector multiplets (\ie complex K\"ahler
moduli) orthogonal to the volume. The field $\phi_4$ is
the four dimensional dilaton and is related to the ten dimensional
dilaton $\phi$ appearing in eq.\dilat\ in a somewhat complicated way owing
to the equations \dilat\ and \Rij. The ten dimensional dilaton kinetic
term when expressed in terms of the 4-dimensional 
dilaton gives extra contribution to the kinetic term for the
scalar field corresponding to the volume. It is for this reason that we
have given the kinetic terms in eq.\IIAmetric\ only for scalars that are
orthogonal to the volume. The kinetic term for the volume, which is a
vector
modulus, can be obtained by using the special geometry in terms of the
prepotential given below.
The metric $G$ above refers to the original
moduli metric for
compactification on the uncorrected Calabi-Yau space. The relative sign between the
K\"ahler and complex structure moduli above is consistent
with the fact that under mirror symmetry $\chi\rightarrow -\chi$ and the K\"ahler
moduli get exchanged with the complex structure moduli. 

The correction to the four dimensional Einstein term however comes entirely from
$Y_2$, and therefore for $IIA$ it is of opposite sign for the tree level and the
one loop:
\eqn\IIAR{ \int d^4x \sqrt{-g} {\chi \over {(2\pi)^3} }
(e^{-2\phi_4}{2\zeta(3)\over {\cal V}} - {2\pi^{2} \over 3})R}
Going now to the Einstein frame one finds the following correction to the
metric of K\"ahler moduli (orthogonal to the volume as described above) 
and of the complex structure moduli:
\eqn\IIAvectorE{ G_{\Lambda \Delta} \rightarrow (1- 4{\chi \over
{(2\pi)^3} } {\zeta(3)
\over {\cal V}}) G_{\Lambda \Delta}}
\eqn\IIAhyperE{ G_{q \bar{q}} \rightarrow (1 + e^{2\phi_4}
{1 \over {6 {\pi}}}\chi) G_{q \bar{q}} }
Noting that the Calabi-Yau volume ${\cal V}$ is part of a vector multiplet
and the four
dimensional dilaton $\phi_4$ is part of a hypermultiplet, we see that the above
corrections are consistent with $N=2$ supersymmetry which dictates a decoupling of
vector and hypermultiplet moduli. The correction to the vector multiplet geometry
appears only at the tree level and in fact corresponds exactly to the four 
loop correction to
the prepotential\foot{In fact the prepotential (2.20) reproduces \IIAvectorE\ 
in the large volume limit for the components of the metric orthogonal to $\cal
V$. Furthermore, consistency of these equations requires an appropriate
normalization of the volume.}
in $IIA$ theory within the CY $\sigma$-model computation 
\refs{\cand,\gvvz}
\eqn\prep{{\cal F}(Z) = {1 \over 6} Z^{\Lambda} Z^{\Delta} Z^{\Sigma}
C_{\Lambda \Delta \Sigma} - i{{\zeta(3) \chi} \over {2 (2 \pi)^3}}
+ \ldots}
On the other hand the correction to the hypermultiplet geometry appears at one-loop
level and is universal. We note that both tree-level and one-loop
corrections
are absent for  CY threefolds with $\chi =0$. This is essential for having a
quantum-exact moduli space and second-quantized mirror symmetry in the $(11,11)$
CY threefold of \fhsv.
The type $IIB$ case will be analysed in Section 5.

We finish this section with
a comment about fourfolds.
The above-mentioned condition for existence of nowhere-vanishing spinors \ipw,
gives rise to tadpole
terms in $D=3$ $N=2$ proportional to the Euler number of the internal
manifold \bb\ that may be cancelled by appropriate membrane configuration
\svw. It is easy to see that in addition to this tadpole, \newrnew\ gives
rise to its supersymmetric extension $\chi \cdot e\wedge e\wedge e$ which
is simply a three-dimensional cosmological constant.

\newsec{One-loop Correction to the Hypermultiplet Geometry}

In this section we will obtain the 1-loop correction to the Einstein term
as well as the metric for hypermultiplets directly from string theory. First let
us consider the correction to the Einstein term. 
This has been earlier computed in
\kkoun\ for the case of orbifold compactification, but here we shall
do it
for the general Calabi-Yau case. We will  obtain the correction to the Einstein term 
by computing 
a 3-point function involving gravitons. From the effective field theory,
this would be the
sum of the irreducible 1-loop 3-graviton vertex plus the diagram corresponding to
an intermediate graviton
connecting the 3-graviton vertex at the tree level
and the 1-loop 2-graviton vertex.
Each of these diagrams is proportional to the 1-loop correction to the Einstein
term, and it is easy to see that the proportionality constant is not zero.

The graviton vertex with the polarization $h_{\mu\nu}$ in the zero ghost 
picture is
\eqn\veras{V_h(p) = h_{\mu\nu} :( \partial X^{\mu}+ip\cdot\psi
\psi^{\mu}) ( {\bar \partial}
X^{\nu} + ip \cdot {\tilde \psi} {\tilde \psi}^{\nu}) e^{i p \cdot X}:\ ,}
where $X^{\mu}$ and $\psi^{\mu}$ and ${\tilde \psi}^{\mu}$ are the bosonic
and the left and right moving fermionic space-time 
coordinates
(in the $NSR$ formalism) and $p$ refers to the 4-dimensional momentum. The
contribution to
the Einstein term being CP-even can come only from spin structures that
correspond to left and the right sectors being both even or odd. Let us first
consider the odd-odd spin structure. In this case, due to the presence of a
holomorphic (anti-holomorphic) killing
spinor, one of the graviton vertex must
appear in (-1,-1) ghost picture: $e^{-\phi -\tilde{\phi}} \psi^{\mu}
\tilde{\psi}^{\nu} e^{ip\cdot X}$ where $\phi$ and $\tilde{\phi}$ are the
bosonization of the left and the right moving superghosts. Moreover, the presence of
the world-sheet gravitino zero modes imply the insertion of a left and right moving
picture changing operator: $P_L = e^{\phi} T_F$ (and similarly the right moving
$P_R$) where $T_F= \psi^{\mu} \partial X^{\mu} +...$ is the superpartner of the
world sheet stress energy tensor. In the odd-odd spin structure, there are four
space-time fermion zero modes each in the left and the right sectors, and 
therefore
one of the graviton vertex in (0,0) ghost picture must give the momentum 
dependent
fermion bilinear piece. The four space-time left moving (and right moving) zero
modes are soaked by the two fermions from this (0,0) ghost-picture graviton and
one each from the (-1,-1) ghost-picture graviton and the $T_F$ in the picture
changing operator. The remaining bosonic part $\partial X$ (${\bar\partial}X$) 
on the
left (right) sector in the $T_F$ and the remaining (0,0) ghost picture graviton, 
will contract with the 
corresponding right (left) moving parts through the propagator
\eqn\veras{\langle  \partial X^{\mu}(z) \bar{\partial}
X^{\nu}(w)\rangle =- {\pi \delta_{\mu\nu} \over {\rm{Im}}\tau}\ ,}
where $\tau$ is the Teichmuller parameter of the world-sheet torus.
After soaking the space-time fermion zero modes, the non-zero mode determinants
of the space-time bosons and fermions cancel. In the internal $N=2$ conformal
field theory describing the Calabi-Yau space, the amplitude is proportional to
the Witten index ${\rm{tr}}(-1)^{F_L+F_R} q^{L_0} \bar{q}^{\bar{L}_0}$, where
$F_L$ and $F_R$ are the left and right moving charges with respect to the 
$U(1)$'s 
of the respective
$N=2$ superconformal algebras, $L_0$ and $\bar{L}_0$ the left and right
moving dimensions, and $q=e^{2i\pi \tau}$.
The Witten Index for the Calabi-Yau space is just its
Euler characteristic $\chi$. This therefore gives an amplitude which is 
quadratic in
momenta with a coefficient proportional to $\chi, \int {d^2\tau
\over {({\rm{Im}}
\tau)^2}}
=2\pi \chi/12$. The dependence on $\rm{Im}\tau$ above follows from the fact that
in four space-time dimensions the partition function comes with $\rm{Im}
\tau ^{-3}$
on the torus, while the two bosonic correlation functions $\langle \partial X
\bar{\partial} X\rangle$ give $\rm{Im}\tau ^{-2}$ and the integrations over the
positions of the three graviton vertices give $\rm{Im}\tau^3$.

Let us now consider the contribution of the even-even spin structures to the
3-graviton amplitude involving two powers of momenta.. In the
even-even spin structure we can use the (0,0) ghost pictures for all the three
gravitons with no insertion of picture changing operators. It is easy to see 
that one needs at least
one fermion propagator from each of the left and the right moving sectors, as 
otherwise the sum over even spin structures would yield a vanishing
result due to $N=2$
space-time supersymmetry. Thus from the left (and right) sector, two of the 
graviton vertices (say at $z_1$ and $z_2$) should provide the fermion bilinear
pieces.
The $\partial X$ part of the third graviton vertex (say at $z_3$) must contract with
$\bar{\partial} X$ part of another graviton vertex (say at $z_1$). Thus the fermion
bilinear pieces from the right moving sectors are provided by the gravitons at $z_2$
and $z_3$. Therefore, the amplitude to begin with is of order $p^4$. The
only way this can contribute to Einstein term, \ie quadratic in momenta,
is if there is a contact term due to singularity in the integral over the
positions of the
graviton vertices. However, 
the correlation function of the fermions in the left sector is 
\eqn\veras{S_e(z_1,z_2)^2
=-\partial_{z_1}^2 \log \theta_1 (z_1-z_2) + 2\pi i\partial_{\tau} \ln({\theta_e
\over
\eta}),}
 where $S_e$ is the Szego kernel in the even spin structure
labelled by $e$, $\theta_1$
and $\theta_e$ are the odd and even Jacobi theta functions, and $\eta$ is the
Dedekind eta function. The right moving part gives a similar correlation function
depending on $(\bar{z_2}-\bar{z_3})$. This shows that there is no singularity of the
form $|z_i-z_j|^2$, and hence  there is no contact term. 
It follows that the even-even spin
structure can only contribute to the $R^2$ term and not to the Einstein term. 
This was indeed expected from the 1-loop 10-dimensinal $R^4$ term which contains
two pieces: one appearing
with the combination $t_8 \cdot t_8$ and the other with $\epsilon \cdot 
\epsilon$, that 
come respectively in the even-even and odd-odd spin structures. The
first term above does not contribute to the Einstein term while the second 
one gives a contribution proportional to the Euler characteristic of the
Calabi-Yau 3-fold. 

Now let us turn to the question of the 1-loop correction to the metric 
of the vector-
and hyper-multiplet moduli. How do we extract this correction from an on-shell
amplitude. The first thing to note is that the 2-point function of the 
corresponding
scalars is zero on-shell.  Moreover the 3-point function involving two 
scalars and a
graviton at one-loop level gets contribution from the irreducible 1-loop
3-point vertex as well as the reducible diagram involving the wave function
renormalization of the scalar with the tree-level 3-point vertex of two 
scalars and a graviton. Each of the above contributions is proportional to the one
loop correction to the
metric and in fact they cancel each other. There is also another reducible
diagram involving the tree-level 3-point vertex of two scalars and a graviton
together with the one-loop graviton self-energy but this depends only on the one
loop correction to $R$ term and does not carry the information of the 1-loop
correction to the metric. Thus, in
order to get the information about the
one-loop correction to the metric, we must compute a 1-loop 4-point amplitude
involving two
scalars and two gravitons. Note that the corresponding string amplitude will 
also include one-loop correction to the Einstein term via one-loop 2 or
3-graviton
vertex with the tree-level kinetic term of the scalars. In fact the sum of 
all these diagrams gives the correct one-loop correction to the metric in
the one-loop
corrected Einstein frame. Thus, a vanishing (non-vanishing) string amplitude 
would imply vanishing (non-vanishing)
one-loop correction to the moduli metric in the Einstein frame, in which the
coefficient of $R$, including the one-loop correction, has been scaled to 
unity. 

We are thus led to computing a 4-point amplitude $\langle hhq\bar{q}\rangle$
where $h$ are graviton vertices and $q$ and $\bar{q}$ are the $NS$-$NS$ scalar 
and its complex conjugate respectively. The vertex operator for the scalar
$q$ is \eqn\versc{V_q(p) = :P_L P_R e^{-\phi-\tilde{\phi}} \Psi e^{i p
\cdot X}:\ ,} where $P_L$ and $P_R$ are the picture changing operators and
$\Psi$ is a chiral-chiral or chiral-antichiral operator for type $IIA$ hypermultiplet
or vector multiplet, respectively, and vice versa for type
$IIB$. The vertex operator for $\bar{q}$ is just the complex conjugate of
the above. 

The amplitude in question is CP-even and therefore gets contribution only from
even-even and odd-odd sectors. Let us first consider the even-even sector. As
mentioned earlier, in order to get a non-vanishing result after the 
spin-structure sum the amplitude must involve fermion correlators from both
the left and the right
sectors. The lowest possible power of momenta is therefore $p^4$, with $p^2$ 
coming from each of the two sectors. In order to obtain one-loop correction to the
metric, which is quadratic in momenta, we must look for possible singularities in
the integrals
over the positions of the vertices which could give a $1/p^2$ pole. Now $p^4$ can
appear in three ways:

a) two gravitons provide $p^2$ from left as well as right sector. In this case 
the correlation function is proportional to $|S_e(z-w)|^4$, $z$ and $w$ being the 
positions of the graviton vertices, which after integration using the formula
(3.3) yields a constant. The scalar operators 
$\Phi= \oint T_F \oint \tilde{T}_F \Psi$ and its
complex conjugate  give rise to the second derivative
$\partial_q \partial_{\bar{q}}$ of the partition function of the internal
conformal field theory in the even-even spin structure and therefore there is no
singularity in the position integrals of the $q$ and $\bar{q}$ vertices. Thus, 
there is no contact term of the form $1/p^2$.

b) two gravitons provide $p^2$ from the left sector and the two scalars provide
$p^2$ from the right. In this case the right moving parts of the graviton
vertices $\bar{\partial} X$ necessarily bring down more powers of momenta.

c) the two scalar vertices provide $p^2$ from both the sectors and the
left moving bosonic parts $\partial X$ of the graviton vertices contract with 
the corresponding right moving parts as in the computation of the one-loop
correction to $R$ above. The correlation function now is
$S_e(z-w)\bar{S}_{e'}(\bar{z}-\bar{w}) \langle 
\Psi(z,\bar{z})
\bar{\Psi}(w,\bar{w})\rangle_{e,e'}$, where $e$ and $e'$ are the even
spin-structures 
on the left and the right sectors correspondingly, and $z$ and $w$ are the
positions of the vertex operators for $q$ and $\bar{q}$, respectively. This
correlation function has the leading singularity structure $1/|z-w|^4$ which is
proportional to the tree-level metric $G^{(0)}_{q\bar{q}}$. Now
the spin structure dependence in the internal conformal field theory 
describing the Calabi-Yau space, enters only through the  charge
lattice of the $U(1)$'s of the left and right moving $N=2$ superconformal 
algebras (see for example \agnta).
Thus, we can extract the spin
structure dependent part of the correlation function for the left moving 
sector with the result:
\eqn\veras{\theta_e(z-w) Z_e(\pm{1 \over \sqrt{3}}(z-w))\ ,}
where $Z_e$ is the
$N=2$ $U(1)$ charge lattice in the same spin-structure and the
$\pm$ sign in the argument of $Z_e$ above is for $q$ being chiral or 
anti-chiral in the left moving sector, following the corresponding $U(1)$
charges $\pm{1\over \sqrt{3}}$. 

One can now carry out the spin structure sum using the formula:
\eqn\veras{\sum_e\theta_e(a) Z_e(b)+ \theta_1(a) Z_1(b) =
2\theta_1 ({{a+\sqrt{3}b} \over 2})
Z_1({{\sqrt{3} a-b}\over 2})\ ,}
where $Z_1$ is the $U(1)$ charge lattice in the odd spin structure. Since
$\theta_1(z-w)$  goes to zero as $z\rightarrow w$, the leading
singularity in the correlation function, including the right moving sector is
$1/|z-w|^2$ which is precisely the singularity needed to produce a $1/p^2$ pole.
This means that in the above equation we can set the argument of $Z_1$ equal to
zero. It then follows from the eq.\veras\ that the contribution of
the sum over the even spin structures to the metric correction is equal to plus 
or minus $\theta_1'(0) Z_1(0)$ for
$q$ being chiral or anti-chiral, respectively.
Note that $\theta_1'(0)$ exactly cancels the contribution of the two 
space-time bosonic non-zero mode determinants.  

In the odd-odd spin structure, as discussed earlier we have to use one of the
vertex operators (say $q$) in (-1,-1) ghost picture and we have to insert also
a picture changing operator each for the left and right moving sector. After soaking
the space-time
fermion zero modes, and contracting the $\partial X$ in one of the
graviton vertex with the right moving part via the correlator (3.2), we
find again a $p^4$ term. The space-time bosonic and fermionic non-zero
mode determinants
cancel as usual. In the internal theory on the other hand there is a correlation
function $\langle \Psi(z,\bar{z}) \bar{\Psi}(w,\bar{w})\rangle$ which has a 
leading singularity
$1/|z-w|^2$ with coefficient $G^{(0)}_{q\bar{q}}|Z_1(0)|^2$, which upon integration
gives rise to a
$1/p^2$ contact term. This result is therefore exactly the one obtained
for the sum
over the even spin structures.

Combining the above contributions and taking into account the relative
sign between the even-even and odd-odd spin structures for type $IIA$ and $IIB$, 
we find that the total result vanishes if $q$ is chiral-antichiral for $IIA$
or chiral-chiral for $IIB$. This means that in both type $II$
theories, the one-loop correction to the metric of vector
multiplets vanishes (in the one-loop corrected Einstein frame). Of course, this is
expected from the non-renormalization of the vector moduli space due to 
the fact that the dilaton in type $II$ theories on Calabi-Yau spaces sits in
hypermultiplet.

On the other hand for hypermultiplets (\ie $q$ being chiral-chiral for $IIA$ or
chiral-antichiral for $IIB$) the even-even sector adds up to the odd-odd sector and 
yields a total of twice the latter. The result as we have seen above,
after extracting the leading singularity, is just the partition 
function  of the internal conformal field theory in the odd-odd spin
structure which is proportional to the Euler characteristic $\chi$ of the
Calabi-Yau space. To summarize, one-loop string computation shows that
while the metric for vector multiplets is not renormalized (as expected),
the metric of hypermultiplets gets a universal one-loop correction proportional
to the Euler characteristic of the Calabi-Yau space times the tree-level 
metric. 

Finally we would like to comment on the implication of the string
computation presented in this section to the kinetic terms in
eq.\IIAmetric. First thing to note is that in the odd-odd spin structure
the contribution comes entirely from the leading singularity in the
$\Psi$-$\bar{\Psi}$ OPE. This corresponds to a factorized diagram
involving
a tree level two scalar and a graviton vertex with the one loop three
graviton vertex, the first being proprtional to the tree level metric of
the scalars while the second one represents the one loop correction to the
Einstein term $R$. Indeed as we had mentioned earlier the correction to
$R$ comes from the $\epsilon \cdot \epsilon$ part of $R^4$ term in
10-dimensions which appears in the odd-odd spin structure. The
even-even spin structure on the other hand gives the $t_8 \cdot t_8$ part of
$R^4$ term and hence does not give correction to the Einstein term.
Therefore the result appearing in the even-even spin structure should 
correspond to the kinetic terms (in the string frame) given in 
eq.\IIAmetric. Indeed this is consistent with the fact that in the string
calculation the leading singularity between the vertex operators for the
two scalars cancels upon summing over the even-even spin structures. What
survives is in fact the subleading singularity which should be interpreted
as the irreducible part of the correction to the kinetic terms. Futhermore
the even-even spin structure contribute to the K\"ahler and complex
structure moduli equal but with opposite sign which is consistent with 
eq.\IIAmetric.

\newsec{Chiral Fivebrane in Type $IIB$ Theory}

In this section, we  show the emergence of a chiral fivebrane
with a (2,0) tensor multiplet supported on its worldvolume in type $IIB$
theory compactified on $S^1$.
This can be predicted on the basis of duality
of this theory with type $IIA$ on $S^1$ by matching the corresponding branes.
The fact that counting matches is well-known \schrev, but the fact that type
$IIA$ theory contains a chiral fivebrane in ten dimensions while type 
$IIB$ does not, suggests that 
the fivebrane which is a magnetic source for the vector $A_{\mu}=g_{\mu
9}$ in type $IIB$ must be chiral. The existence of this third fivebrane is also
in agreement with the $IIB$ supersymmetry algebra in ten dimensions which has
a triplet of self-dual fivebrane central charges \townss\foot{A similar
mechanism can be seen when deriving the five-dimensional $N=8$ supersymmetry algebra
from eleven dimensional superalgebra with two- and fivebrane central charges. In
addition to 27 (scalar) electric and 27 (vector) magnetic charges, an 
$E_6$ singlet emerges which has the interpretation of the KK monopole charge
\bars. When the theory is reduced to four dimensions, these charges together
with the extra KK charge complete the 56-dimensional representation of $E_7$.}.

A direct zero mode analysis can be
done following \chs, but here we present only a simple argument based on
worldvolume anomalies. 
In eleven dimensions, it was pointed out \refs{\town,\witfb}\
that, in the presence of a fivebrane, a
term representing the coupling of an anti-self-dual three-form
field strength $T_3$ on the fivebrane worldvolume is necessary to cancel 
the anomaly from the interaction $\int_{M^{11}} A_3\wedge F_4 \wedge F_4$. 
This can be seen as follows. In the presence of a fivebrane with charge $m$,
\eqn\fbsource{dF_4=m\delta_V,}
where $\delta_V$ is supported on the fivebrane worldvolume $V$
({\it i.e.} it integrates to $1$ on the space transverse to the fivebrane).
So, under 
$\delta A_3=d\Lambda_2$,
\eqn\varbulk{\eqalign{{1\over 12}
\delta\left(\int_{M^{11}} A_3\wedge F_4 \wedge F_4\right)
&={1\over 4}\int_{M^{11}} d\Lambda_2\wedge F_4 \wedge F_4\cr
&=-{m\over 2} \int_V \Lambda_2\wedge F_4.\cr}}
This anomaly needs to be cancelled by a term
\eqn\sdtensor{{m\over 2}\int_V T_3\wedge A_3,}
where $T_3$ is the anti-self-dual three-form field 
strength on the fivebrane worldvolume and $dT_3=F_4$. 

Similarly, the kinetic term for the self-dual four-form field in type
$IIB$ yields in nine dimensions, besides the kinetic term of the
three-form field, a Chern-Simons (CS) coupling involving the vector
$A_{\mu}$: $\int_{M^{9}} A_1\wedge F_4 \wedge F_4$. Here, we don't
consider other couplings, involving $A_3$ with Neveu Schwarz ($NS$) 
and Ramond-Ramond $(RR)$
two-form fields  inherited from ten dimensions since they are not relevant for
our analysis. The fivebrane  source with charge $m$ is now,
\eqn\fbsourcenew{dF_2 = m\delta_V,}
and  under $\delta A_3 = d \Lambda_2$, the CS coupling develops exactly the
same anomaly as before which can be remedied only by introducing a
coupling to a tensor on the worldvolume \sdtensor. Again, this is an
indirect argument which can be
confirmed by the zero-mode analysis. Similar analysis in type $IIA$
shows that the fivebrane charged under $A_{\mu}=g_{\mu 9}$ is
non-chiral. Thus in nine dimensions, the quantum consistency of the theory
requires a coupling involving an eight-form in curvature that is formally
given by the gravitational anomaly of a six-dimensional $(2, 0)$ tensor
multiplet:
\eqn\ninecop{\int A_1 \wedge I_8(R).}

Note that under the $U$-duality group $SL(2, Z)$ the three nine-dimensional 
vectors form a doublet (non-chiral fivebrane) and a
singlet (chiral fiverane). While the singlet in type $IIB$ is the
$A_{\mu}=g_{\mu 9}$ vector, in type $IIA$ it is the $A_{\mu} = B_{\mu 9}$
which is consistent with the fact that the $T$-duality connecting the
two theories in nine dimensions  interchanges the internal components of
the metric and antisymmetric tensor.
Thus we have shown the existence of  a supersymmetric partner of the
$R^4$ terms in nine dimensions.

A similar mechanism arises in the reduction of six-dimensional chiral
theories to five dimensions. A theory with $n_T$
tensor multiplets and $n_V$ vector multiplets have gravitational
Chern-Simons couplings (only the
scalars belonging to the tensor multiplets couple to $R^2$ terms). However, 
the five-dimensional theory contains $n_T + n_V + 1$ vectors and,
generally speaking, all of them have gravitational couplings. In 
particular, we
observe again that the extra vector multiplet coming from the metric
acquires a
gravitational coupling due to the fact that the BPS string, which is
magnetically charged under this multiplet, is chiral.
 
\newsec{Universal Corrections to $IIB$ Theory}

In this section, we discuss modifications of the two-derivative terms in the
effective action of type $IIB$ compactifications
on CY threefolds to four dimensions.
In ten dimensions, the Einstein term together with the $R^4$ action \newrnew\
can be written, in the string frame, as \gmg:
\eqn\gmgten{\sim\int_{M_{10}}\sqrt{-g} \left[ e^{-2\phi_{10}} R +  
e^{-\phi_{10}}f(\rho, {\bar \rho}) Y^{IIB} \right]  ,}
where $\rho=\rho_1+i\rho_2$ with $\rho_1$ the $RR$ scalar and $\rho_2 =
e^{-\phi_{10}}$. The function 
$Y^{IIB} = {1 \over 3 \cdot 2^8}(t_8 \cdot t_8 R^4
+{1 \over 4} \epsilon \cdot \epsilon R^4) \equiv Y_0 +Y_2$ (in the notation
of Section 2), where
we have used the relative sign that appears at the tree level of $IIA$ theory.
Indeed at the tree level there is no difference in $R^4$ term for $IIA$ and 
$IIB$ theory. The $SL(2,Z)$ invariance of $IIB$ then fixes the sign of $R^4$ 
terms for all perturbative and non-perturbative corrections. This of course 
implies that one-loop $\epsilon \cdot \epsilon$ terms for $IIA$ and $IIB$ 
have opposite signs. We shall also see below that the absence of 
hypermultiplet dependence of
the four dimensional $R^2$ term (which is dictated by supersymmetry) implies
precisely this choice of relative sign. The function $f$ is $SL(2,Z)$ 
modular invariant and its explicit form has been conjectured in ref. \gmg\ : 
\eqn\expff{f(\rho, {\bar \rho}) =
{\sum}'_{n,m\in Z}{\rho_2^{3/2}\over |n+m\rho |^3}
= 2 \zeta(3) \rho_2^{3/2} + 
{2 \pi^2 \over 3} \rho_2^{-1/2} + \ldots}
The first two terms in the r.h.s. of \expff\ correspond to 
the tree-level and one-loop
contributions while the dots stand for the instanton sum.

Upon compactification to four dimensions and using \twor, one finds
\eqn\gmgterm{\int_{M_4} \left[ e^{-2\phi_4} + {{\chi} \over {(2 \pi)^3}}  
( 2 \zeta(3) {e^{-2\phi_4} \over {\cal V}} + {2\pi^{2} \over 3} + 
\ldots) \right] \sqrt{-g} R ,}
where $\phi_4$ is the four-dimensional dilaton, $e^{-2\phi_4} = e^{-2\phi_{10}} 
{\cal V}$ plus corrections owing to equations similar to  \Rij and \dilat.
Here $\cal V$ is the CY volume. Note that the relative sign
between the tree
level and the one-loop term above is opposite to that of type $IIA$
\IIAR. On the
other hand since the moduli metric gets contribution only from the $Y_0$ term
above it
should have the same form as in \IIAmetric\ for the $IIA$ case:
\eqn\IIBmetric{ \int d^4x \sqrt{-g} {\chi\over {(2\pi)^3}}
(e^{-2\phi_4}
{2\zeta(3) \over {\cal V}} + {2 \pi^2 \over 3} + 
\ldots) (G_{\Lambda \Delta}\partial
Z^{\Lambda} \partial
{\bar{Z}}^{\Delta}-  G_{q\bar{q}}\partial q \partial{\bar{q}})}
Here we have flipped the overall sign as compared to \IIAmetric\ due to the fact
that in $IIB$ K\"ahler and complex structure moduli correspond to hyper and vector
multiplets as opposed to the $IIA$ case where they correspond to vector and 
hypers respectively \foot{Note that in terms of K\"ahler and complex structure moduli,
\IIBmetric\ and \IIAmetric\ are identical. Throughout the paper, we denote 
the vector and hyper moduli as $Z^{\Lambda}$ and $q$ respectively.}.
Just as in \IIAmetric, the above formula is true only for K\"ahler moduli
(which are hypers in the present case) that are orthogonal to the volume.

By redefining the dilaton as in \couplhype\ and going to the
Einstein frame one finds the following corrections to the metric for 
the vector moduli and that for the hyper moduli that are orthogonal to
the volume: 
\eqn\IIBvectorE{ G_{\Lambda \Delta} \rightarrow  G_{\Lambda
\Delta}}
\eqn\IIBhyperE{ G_{q \bar{q}} \rightarrow [1 - {2\chi\over {(2\pi)^3}}
({2\zeta(3) \over {\cal V}} + e^{2\phi_4}{2 \pi^2 \over 3} + 
\ldots)] G_{q \bar{q}} }

Thus the metric of the vector multiplet is not corrected even at tree
level.
On the other hand  eq.\IIBhyperE\ implies
a universal correction to the
hypermultiplet quaternionic geometry, in the large volume limit. In fact, 
the second (one-loop) term was already obtained in Section 3 by a direct
string calculation. 
The tree-level term $\sim
{\zeta(3)\chi}$ is the contribution which should come from the classical $c$-map
\refs{\cfg,\fsab}, due to the modification of the prepotential from the
zero-instanton sector. In other words, this modification can
be identified  on the type $IIA$
side as a four-loop contribution to the prepotential within the CY
$\sigma$-model
computation and was discussed in Section 2 (see \prep). Note that although
the tree level correction implied in \IIBhyperE\ is given here only for
the hypers that are orthogonal to the volume, it can be extended to
include volume by using the tree level special geometry for the NS-NS part
of the non-universal hypers. 
{}From this argument we learn that type $IIB$ hypermultiplet geometry, in the
large volume limit, receives tree-level, one-loop and non-perturbative
corrections
\foot{Note that only the tree-level term is consistent
with the classical $c$-map since it can be reabsorbed in a
modification of the prepotential.}. Again, such corrections are absent for CY's
with $\chi =0$ \fhsv.

One type of universal non-perturbative corrections to hypermultiplets
has been discussed in \becker\ and involves fivebrane instantons. In type
$IIB$ theory
with a non-vanishing $RR$ fields, there can be two euclidean fivebranes
wrapped around the CY space, and thus one expects that the results of
\becker\ are modified. It is known that the two dilatons and two
antisymmetric tensors form the universal hypermultiplet and their
classical moduli space is locally
$SU(2,1) \over SU(2) \times U(1)$. The effect of the wrapped instantons
should be combined together with the instanton corrections present
already in ten dimensions, and as a result the whole system should respect
the quaternionic structure. Indeed, this multiplet is not sensitive to 
the internal space, and  thus its monodromy seems to be a
direct generalization of $SL(2,Z)$ and is of purely gravitational nature.
It is suggestive that $SL(2,Z)$ is enlarged to $SU(2,1,Z)$, and leads us
to a speculation  that in four
dimensions the $SL(2, Z)$ modular function $f$ conjectured in \gmg\ should
be promoted to a quaternion-valued function ${\hat f}(Q)$, 
where $Q$ is built from the universal 
hypermultiplet\foot{${\hat f}(Q)$ should have the basic
properties of $f(\rho, \bar{\rho})$ which should be reproduced in
an appropriate (decompactification) limit when the fivebrane instantons
are suppressed.}. Another argument in favour of $SU(2,1,Z)$ is that 
(at least on a manifold mirror to one with $h_{(2,1)}=0$ \cdp) the quaternionic
geometry is 
likely to  have both ten-dimensional $SL(2,Z)$ $U$-duality  as well as 
four-dimensional $SL(2,Z)$ $S$-duality, with a common $Z_2$ symmetry that 
inverts the coupling
constant but two  different Peccei-Quinn symmetries (one $RR$, one $NS$-$NS$).
In a sense, this is a complimentary case to the one studied in \ooguri\
where $h_{(2,1)}$ was taken to be one and the universal multiplet was frozen.
One should note though that in our case the hyper-K\"{a}hler limit is trivial. 

\newsec{Concluding Remarks}

After having explained in Section 2 the K\"ahler class dependence of $R^2$ terms 
for a generic CY compactification of type $IIA$ theory as a consequence of $R^4$
coupling, we address now the question of the origin of additional terms 
required by spacetime supersymmetry. 
Recall that in four dimensions $R^2$ couplings come from a 
supersymmetric action of the type 
${\cal F}_1W^2|_{\rm F-term}$,
where $W$ is the $N=2$ chiral Weyl superfield and ${\cal F}_1$ 
is a function of chiral 
vector superfields \agnta. Expanding in components, one finds that in addition
to $R^2$ there are other four-derivative terms of the form $T^2F^2$ and $TFR$,
where $T$ and $F$ are the field-strengths of the $N=2$ graviphoton and ``matter" 
vectors, respectively. The graviphoton originates from the eleven-dimensional
metric while the gauge fields come from the three-form $A_3$. Since in the large
volume limit, the function of the moduli $f(t)=\alpha_{\Lambda} t^{\Lambda}$
coupled to $R^2$ is linear, the $T^2F^2$ term drops out. Moreover, since 
$\alpha_{\Lambda}$ are real, only the imaginary part of $FTR$ term in the
superfield expansion will contribute. This term comes from the original $A_3
\wedge X_8(R)$ term. To see this lets integrate \new\ by part to get 
$\alpha_{\Lambda} F_2^{\Lambda} \wedge \omega_3$ and note that upon reduction
on $S^1$, this contains
\eqn\zzz{\alpha_{\Lambda} F_2^{\Lambda} \wedge RT,}
where two-form $RT$ stands for $R_{\mu \nu}^{\sigma \lambda} T_{\sigma
\lambda}$.

Turning to type $IIB$ theory, we first note as a consistency check that 
according to special geometry there should be no loop corrections associated 
with the K\"{a}hler class moduli. Indeed, with the different choice of relative
sign in  \newrnew\ the  two  contributions to the four-dimensional 
$(c_2 \cdot{\vec J}) R^2$  term cancel and there are no such corrections. 
Similar cancellation holds for the $R^2$ couplings in $D=6$ when type $IIB$ 
is compactified on $K_3$.

For the discussion of complex structure dependence of $R^2$ term, we first turn 
to $N=4$ case. The reduction of type $IIA$ on $K3 \times T^2$ gives a term 
$t R^2$, where $t$ is the K\"ahler modulus of $T^2$, and by duality we expect 
a similar
term in type $IIB$ where $t$ modulus is replaced by the complex structure $U$.
The answer is rather obvious if one first compactifies on $S^1$. Then the
anomalous coupling associated with chiral fivebrane discussed in Section 4 
will be
further reduced to the term of the right form. Note that if one now takes for
example a CY that has an orbifold limit $K3 \times T^2/Z_2$, the off-diagonal 
component of the metric on $T^2$ ({\it i.e.} the complex structure $U$)
survives the orbifolding, and one again gets an $R^2$ coupling depending on the
complex structure.

\vskip.5truecm
\noindent
{\it Note added:} The recent paper hep-th/9706195 by A. Strominger 
overlaps  with some of our results, in particular concerning the one-loop 
correction to the universal hypermultiplet.

\vskip1truecm
\noindent
{\bf Acknowledgements:}
\noindent
We would like to thank Elias Kiritsis and Stefan Theisen for helpful
discussions.
This work is supported in  part by DOE grant DE-FG03-91ER40662, Task C., and
in part by the European Commission under the TMR contracts ERBFMRX-CT96-0045 
(LNF, Frascati, INFN)
and ERBFMRX-CT96-0090. 
S.F. would like to thank  Physics Departments of Rutgers University,
the Rockefeller University and NYU were part of this work was done.
R.M. is supported by a World 
Laboratory Fellowship.

\vfil\eject
\listrefs
\bye